\documentclass[prl,aps,twocolumn,floatfix,amsmath,notitlepage]{revtex4-1}

\usepackage{amssymb}
\usepackage{graphicx}
\usepackage{graphics}
\usepackage{amsmath}
\usepackage{amsthm}
\usepackage{color}
\usepackage{dsfont}
\usepackage{mathrsfs}
\usepackage{float}

\usepackage{mathtools}
\usepackage[unicode=true,pdfusetitle,bookmarks=true,bookmarksnumbered=false,bookmarksopen=false,breaklinks=false,pdfborder={0 0 0},pdfborderstyle={},backref=false,colorlinks=true]
{hyperref}

\def\>{\rangle}
\def\<{\langle}

\renewcommand{\qedsymbol}{\nobreak \ifvmode \relax \else
	\ifdim \lastskip<1.5em \hskip-\lastskip \hskip1.5em plus0em
	minus0.5em \fi \nobreak \vrule height0.75em width0.5em
	depth0.25em\fi}

\renewcommand{\geq}{\geqslant}
\renewcommand{\leq}{\leqslant}

\def\p{\mathbf{p}}
\def\rr{\mathbf{r}}
\def\qq{\mathbf{q}}

\newtheorem{theorem}{Theorem}
\newtheorem*{theorem*}{Theorem}

\newtheorem*{lemma*}{Lemma}

\newtheorem{definition}{Definition}
\newtheorem*{definition*}{Definition}

\theoremstyle{remark}
\newtheorem*{remark}{Remark}

\theoremstyle{definition}

\newcommand{\bea}{\begin{eqnarray}}
\newcommand{\eea}{\end{eqnarray}}
\newcommand{\be}{\begin{equation}}
\newcommand{\ee}{\end{equation}}
\newcommand{\ba}{\begin{equation}\begin{aligned}}
\newcommand{\ea}{\end{aligned}\end{equation}}

\newtheorem{example}{Example}

\def\be{\begin{equation}}
\def\ee{\end{equation}}

\newcommand{\lr}{\rangle\langle}

\newcommand{\tr}{{\rm Tr}}

\newcommand{\ve}[1]{{\left\vert\kern-0.25ex\left\vert\kern-0.25ex\left\vert #1 
    \right\vert\kern-0.25ex\right\vert\kern-0.25ex\right\vert}}

\newcommand{\mbb}[1]{\mathbb{#1}}

\newcommand{\bra}[1]{\langle #1|}
\newcommand{\ket}[1]{|#1\rangle}

\newcommand{\eqdef}{\coloneqq}

\usepackage[most]{tcolorbox}
\newtcolorbox{myt}[2][]{%
  attach boxed title to top center
               = {yshift=-4pt},
  colback      = blue!5!white,
  colframe     = blue!75!black,
  halign       = flush left,
  fonttitle    = \bfseries\sffamily,
  colbacktitle = blue!65!black,
  title        = #2,#1,
  enhanced,
}
\newtcolorbox{myd}[2][]{%
  attach boxed title to top center
               = {yshift=-4pt},
  colback      = violet!5!white,
  colframe     = violet!75!black,
  halign       = flush left,
  fonttitle    = \bfseries\sffamily,
  colbacktitle = violet!65!black,
  title        = #2,#1,
  enhanced,
}
\newtcolorbox{mye}[2][]{%
  attach boxed title to top center
               = {yshift=-4pt},
  colback      = purple!5!white,
  colframe     = purple!75!black,
  halign       = flush left,
  fonttitle    = \bfseries\sffamily,
  colbacktitle = purple!65!black,
  title        = #2,#1,
  enhanced,
}

\newtcolorbox{myg}[2][]{%
  attach boxed title to top center
               = {yshift=-4pt},
  colback      = green!5!white,
  colframe     = green!75!black,
  halign       = flush left,
  fonttitle    = \bfseries\sffamily,
  colbacktitle = green!65!black,
  title        = #2,#1,
  enhanced,
}

\begin{document}
	
	%\SetWatermarkText{NOT FOR DISTRIBUTION}
	%\SetWatermarkAngle{60}
	%\SetWatermarkScale{0.5}
	
	\title{What are the minimal conditions required to define a SIC POVM?}

\author{Isabelle Jianing Geng}\email{jianing.geng@ucalgary.ca}\author{Kimberly Golubeva}\email{kimberly.golubeva@ucalgary.ca}\author{Gilad Gour}\email{gour@ucalgary.ca}
\affiliation{
Department of Mathematics and Statistics, Institute for Quantum Science and Technology,
University of Calgary, AB, Canada T2N 1N4}

	\date{\today}
	
	\begin{abstract}
	 Symmetric informationally complete (SIC) POVMs are a class of quantum measurements which, in addition to being informationally complete, satisfy three conditions: 1) every POVM element is rank one, 2) the Hilbert-Schmidt inner product between any two distinct elements is constant, and 3) the trace of each element is constant. The third condition is often overlooked, since it may give the impression that it follows trivially from the second. We show that this condition cannot be removed, as it leads to two distinct values
for the trace of an element of the POVM. This observation has led us to define a broader class of measurements which we call semi-SIC POVMs. In dimension two we show that semi-SIC POVMs exist, and we construct the entire family. In higher dimensions, we characterize key properties and applications of semi-SIC POVMs, and note that the proof of their existence remains open.
\end{abstract}

	\maketitle

	%First paragraph: Introduction
	
    {\it Introduction.} Symmetric Informationally Complete Positive Operator Valued Measures (SIC POVMs) are objects which straddle the junction between mathematics and physics. This particular type of quantum measurement has recently received a great deal of attention in both communities because of its vast array of diverse applications~\cite{Fuchs_2017, Appleby_2017,Bengtsson_2020,Appleby_2020, Durt_2008,Medendorp_2011,Pimenta_2013,PhysRevA.74.042341, fuchs2003squeezing,PhysRevA.70.052321,englert2004efficient,Scott_2006,Lai2020DetectingES,Rastegin2015OnTB,Graydon_2016,Tabia_2013,Zhu_2016,fuchs2010qbism, Appleby_2017, fuchs2016qbism, Fuchs_2013, TBGR2020, BBCE2009, HCM2006}.  SICs are connected to several open problems within the field of algebraic number theory, including Hilbert's 12th problem~\cite{Appleby_2017,Bengtsson_2020,Appleby_2020}. Within physics, SICs are an optimal type of quantum measurement which have been realized experimentally~\cite{Durt_2008,Medendorp_2011,Pimenta_2013,PhysRevA.74.042341}, utilized in quantum information theory~\cite{fuchs2003squeezing,PhysRevA.70.052321,englert2004efficient,Scott_2006,Lai2020DetectingES,Rastegin2015OnTB,Graydon_2016,Tabia_2013,Zhu_2016,fuchs2010qbism} and influenced the foundations of quantum mechanical theory~\cite{Appleby_2017, fuchs2016qbism, Fuchs_2013}. Despite the rapid growth of interest in these objects, a proof of their existence in all finite dimensions--a conjecture first postulated over two decades ago by Zauner--~\cite{ZaunerGerhard} remains elusive. Exact solutions have been found in dimensions $2-24,28,30,31,35,37,39,43,48, 124$~\cite{Renes_2004, Appleby_2019, Appleby_2018, HS2016}and numerical solutions have been found in dimensions $1-151$~\cite{scott2017sics, Scott_2010}, as well as in several other dimensions up to dimension $844$~\cite{Grassl_2017}. 
   
   \emph{Informationally Complete} (IC) POVMs posses the characteristic that, when acting on a particular state, their statistics completely determine the quantum state. More precisely, an IC POVM is described by $d^2$ positive semi-definite operators, $\{E_x\}_{x=1}^{d^2}$, that  span the $d^2$-dimensional space of observables on a $d$-dimensional Hilbert space $\mathcal{H}$. 
\begin{definition}\label{def1}
An IC POVM $\{E_x\}_{x=1}^{d^2}$ is a \emph{symmetric} IC-POVM (in short SIC-POVM) if it satisfies three conditions:
\begin{enumerate}
    \item $E_x$ is rank one for all $x \in  \{1, ... , d^2\},$
    \item The Symmetry Condition; $$\tr[E_xE_y] = b\quad \text{for all } x \neq y\;,$$
    \item $\tr[E_x]= a$ for all $x \in  \{1, ... , d^2\}.$
\end{enumerate}
where $b$ and $a$ are constants and $d$ is the dimension of the underlying Hilbert space. 
\end{definition}

 As aforementioned, in some instances a SIC POVM is an \emph{optimal} type of measurement; a consequence which arises from the property that a SIC POVM is comprised of rank-one operators~\cite{Renes_2004}. In~\cite{REK2004, EKNCRA2004, WF1988}, \emph{optimality} was studied in the context of quantum tomography. More specifically, optimality refers to the minimal error in state estimation. This concept has proven to be of significance in experimental realizations of quantum tomography~\cite{PhysRevX.5.041006}. In addition, it has been shown that when the rank one condition is relaxed, SIC POVMs exist in all dimensions~\cite{Appleby2007, GA2014}, but in this case optimality is lost. The term \emph{symmetric} pertains to their characterization as equiangular tight frames~\cite{Renes_2004} which form the vertices of a regular simplex in a space that contains the convex combinations of quantum states~\cite{Appleby2011}. This condition cannot be relaxed as it is the integral defining characteristic of a SIC POVM. 

The value $a= \frac1d$ is derived from the fact that any POVM satisfies $ \sum_{x=1}^{d^2}E_x=I_d$, where $I_d$ is the identity operator. From here, it follows that we have $\sum_{x=1}^{d^2}\tr[E_x]=\tr[I_d]=d$, which implies that $a=\tr[E_x]=\frac{1}{d}$. 
This third condition is somewhat overlooked since it may give the impression that it follows directly from the second condition. That is, that the value for the trace of an individual element of a SIC POVM follows trivially from the constant-valued Hilbert-Schmidt inner product between any two elements of the SIC. 

In this paper we show that, in the two dimensional case, it is necessary to specify that the trace of an individual element of a SIC POVM has a constant value. More specifically, we conclude that rather than implying  $\tr[E_x]=\frac1d$ for all $x$, the symmetry condition in Definition~\ref{def1} implies that $\tr[E_x]$ can take at most two distinct values. We use the consequences of this result to define a new class of POVMs which we refer to as \emph{semi-SIC POVMs.} After constructing the entire family of semi-SIC POVMs in the two dimensional case, we describe a few key properties of semi-SIC POVMs in arbitrary finite dimensions. Finally, as an application, we calculate the dual basis which enables us to represent a quantum state in terms of a probability vector, analogous to the procedure used with SIC POVMs (e.g.\ ~\cite{Appleby2011}).

{\it Semi-SIC POVM.\;}We begin with the definition of a semi-SIC POVM.
\begin{myt}{{\color{yellow} semi-SIC POVM}}
\begin{definition}\label{main}
Let $\{E_x\}_{x=1}^{d^2}$ be an IC POVM acting on a Hilbert space of dimension $d$. Then $\{E_x\}_{x=1}^{d^2}$ is called a \emph{semi-SIC POVM} if it satisfies the following conditions: 
\begin{enumerate}
    \item $E_x$ is rank one for all $x \in  \{1, ... , d^2\},$
    \item $\tr[E_xE_y] = b$ for all $x \neq y$,
\end{enumerate}
where $b$ is a constant.  
\end{definition}
\end{myt}

We first show that the trace of each element of a semi-SIC POVM can take at most two distinct values.
 Denote Tr$[E_x]:= a_x$ so that each element of a semi-SIC POVM can be written as $E_x = a_{x} \ket{\psi_x}\bra{\psi_x}$. Then,  
\ba
a_{x} &= \tr[E_x] = \tr[E_xI_d]\\
&= \sum_{y\neq x} \tr[E_xE_y] + \tr[E_x^2]\\
&=(d^2-1)b+a_x^2\;.
\ea
This gives us the quadratic equation 
\be
a_{x}^2 - a_{x} + (d^2-1)b = 0\;,
\ee 
which yields two possible values for $\tr[E_x]$. Namely, we have that $a_x\in\{a_+,a_-\}$ where
\be
a_{\pm} \eqdef \frac{1 \pm \sqrt{1-4b(d^2-1)}}{2}\;, \label{eq:1}
\ee
and necessarily $b \leq \frac{1}{4(d^2-1)}$. We will soon see that in dimensions $d\geq 3$, the parameter $b$ can take only a few discrete values, whereas in dimension two, $b$ can take a continuous range of values.

The two distinct traces possessed by the elements $E_x$ prompts us to introduce a new parameter $k$, which will help us to determine the possible $b$ values of a semi-SIC POVM. The parameter $k$  denotes the number of operators in the semi-SIC POVM, $\{E_x\}_{x=1}^{d^2}$, with trace $\tr[E_x]=a_-$. That is, there are $k$ operators with trace $a_-$ and $d^2-k$ operators with trace $a_+$. Since $\{E_x\}_{x=1}^{d^2}$ is a POVM, the trace of all elements must sum to $d$. Hence, 
\be
d = ka_- + (d^2-k)a_+\label{eq:1.1}\; .
\ee 
Substituting the expressions for $a_+$ and $a_-$ from \eqref{eq:1} into equation \eqref{eq:1.1}, we have
\be
d^2-2d=(2k-d^2)\sqrt{1-4b(d^2-1)}\label{eq:1.2}\;.
\ee

Notice that, when $d=2$ the left-hand-side of \eqref{eq:1.2} equals $0$. That is, when $d=2$, equation \eqref{eq:1.2} simplifies to

\be
0 = (2k-4)\sqrt{1-12b}\label{eq:2} \;,
\ee 
which implies that there are only two possibilities; namely, either $k= 2$ or $b=\frac{1}{12}$. However, the latter implies that $a_+=a_-$ which means that the semi-SIC POVM is, in fact, a SIC POVM. Hence, the possibility that does not result in a SIC POVM is $k=2$.

For dimension $d \geq 3$, the left-hand-side of equation~\eqref{eq:1.2} is equal to a positive integer. Hence, for any finite dimension $d \geq 3$ the value of $b$ is given by
\be
    b = \frac{(k-d)(k+d-d^2)}{(d^2-1)(d^2-2k)^2} \label{eq:3} \;.
\ee
From~\eqref{eq:2} it follows that $k$ must be no smaller than $\frac{d^2}{2}$. However, since $b>0$,  the equation above implies that $k$ is bounded by
\be
d^2-d < k \leq d^2\label{eq:4}\; ,
\ee
where the upper bound follows trivially by definition. 
A SIC POVM corresponds to the the case  $k = d^2$ (i.e. $\tr[E_x] = a_-$ for all $x \in  \{1, ... , d^2\}$) which gives the value $b = \frac{1}{d^2(d+1)}$ in \eqref{eq:3}. 

{\it Construction.\;} In the two dimensional (or \emph{qubit}) case, we are able to construct all semi-SIC POVMs up to unitary equivalence. We discover that all semi-SIC POVMs can be characterized in terms of a continuous variable $b$. In the following we construct all semi-SIC POVMs in dimension two.

\begin{theorem}~\label{thm}
Let $b \in \left(\frac{1}{16},\frac{1}{12}\right]$, and define 
\begin{align}
    E_1&\eqdef a_-|\psi_1\rangle\langle\psi_1|\quad,\quad E_3\eqdef a_+|\psi_3\rangle\langle\psi_3|\\
    E_2&\eqdef a_-|\psi_2\rangle\langle\psi_2|\quad,\quad E_4\eqdef a_+|\psi_4\rangle\langle\psi_4|
\end{align}
with $a_{\pm}$ given in~\eqref{eq:1}, and the 2-dimensional vectors $\{\ket{\psi_{x}}\}$ given by
\begin{align*}
    |\psi_1\rangle&\eqdef |0\rangle ,\;\;\;\;\quad\quad\quad\quad\quad\quad |\psi_3\rangle\eqdef \frac{1}{\sqrt{3}}|0\rangle-\sqrt{\frac{2}{3}}e^{i\theta}|1\rangle\\
    |\psi_2\rangle&\eqdef r|0\rangle+\sqrt{1-r^2}|1\rangle ,\;\;\; |\psi_4\rangle\eqdef \frac{1}{\sqrt{3}}|0\rangle-\sqrt{\frac{2}{3}}e^{-i\theta}|1\rangle 
\end{align*}
where  
\be
r\eqdef \frac{2 \sqrt{b}}{1 - \sqrt{1-12b}},\; 
\theta \eqdef \cos^{-1} \left(\frac{\sqrt{1-8b-\sqrt{1-12b}}}{4\sqrt{b}}\right)\;.\nonumber
\ee
Then, $\{E_x\}_{x=1}^{4}$ is a semi-SIC POVM, and for any other semi-SIC POVM in dimension two, $\{G_x\}_{x=1}^4$, there exists a $2\times 2$ unitary matrix $U$ such that $\{UG_xU^{\dag}\}_{x=1}^{4}$ is a semi-SIC POVM of the above form.
\end{theorem}
\begin{remark}
The POVM constructed above is semi-SIC for all $b\in\left(\frac{1}{16},\frac1{12}\right]$, where the case $b=\frac1{12}$ corresponds to a SIC POVM. Since $b\in\left(\frac{1}{16},\frac1{12}\right]$ we must have $r\in\left[\frac1{\sqrt{3}},1\right)$ and $\theta\in\left(\frac\pi3,\frac\pi2\right]$.
\end{remark}
\begin{proof} 
It is straightforward to verify that $\{E_x\}_{x=1}^{4}$ is a semi-SIC POVM. 
We therefore prove that all semi-SIC POVMs in dimension two must take this form. Let $\{G_x=a_x|\psi_x\lr \psi_x|\}_{x=1}^{4}$ be a semi-SIC POVM with $a_1=a_2=a_-$ and $a_3=a_4=a_+$ (recall from the argument below~\eqref{eq:2} that $k=2$). By applying the unitary equivalence, we can assume w.l.o.g. that 
\begin{align}
\ket{\psi_1} &= \ket{0}\;\;,\;\; \ket{\psi_2} = r \ket{0} + \sqrt{1-r^2} \;  \ket{1}\;\;\text{and }\\
\ket{\psi_j} &= s_j \ket{0} + \sqrt{1-s_j^2} \; e^{i\theta_j} \ket{1}\quad\text{for }j=3,4\;,
\end{align}
where $s_j,r\in[0,1]$ and $\theta_j\in\mbb{R}$ for $j=3,4$. The condition $\tr[G_1G_2]=b$ gives
\be
r=\frac{2 \sqrt{b}}{1 - \sqrt{1-12b}} \;. \label{eq:4.1}
\ee
The condition $\tr[G_1G_x]=b$ for $x=3,4$ gives 
\be
s_3=s_4=\frac{1}{\sqrt{3}}\;.
\ee
The condition $\tr[G_2G_3]=b$, gives
\be
\cos(\theta_3) = \frac{\sqrt{1-r^2}}{2\sqrt{2}r}=\frac{\sqrt{1-8b-\sqrt{1-12b}}}{4\sqrt{b}}\;.
\ee
Additionally, $\tr[G_2G_4]=b$ reveals that $\theta_3 = -\theta_4$. \\

It follows from~\eqref{eq:4.1}   that $r\leq1$ corresponds to the condition $b \geq \frac1{16}$, and $r\geq \frac1{\sqrt{3}}$ corresponds to the condition $b \leq \frac1{12}$. The value $b=\frac{1}{16}$ is not permitted since it yields $E_1=E_2$ and $E_3=E_4$ and therefore is not informationally complete.
Moreover, $b$ cannot exceed $\frac1{12}$ since $a_{\pm}$ in~\eqref{eq:1} are not complex. That is,  $b\in\left(\frac{1}{16},\frac{1}{12}\right]$.
\end{proof}

%%%%%%%%%%%%%%%%%%%%%%%%%%%%%%%%%%%%%%%%%%%%%%%%%%

To conclude, we construct an explicit example of a semi-SIC POVM in the qubit case.
\begin{example}\label{example1}
 Let $b=\frac{2}{25}$. Then the four elements of the corresponding semi-SIC POVM are
\begin{align}\label{z1}
    E_1&=\frac{2}{5}
       \begin{pmatrix}
       1&0\\
       0&0
       \end{pmatrix} \;,\;
       E_3=\frac{1}{5}
       \begin{pmatrix}
       1&-\sqrt{2}e^{-i\theta}\\
       -\sqrt{2}e^{i\theta}&2
       \end{pmatrix}\;, \\
       E_2&=\frac{1}{5}
       \begin{pmatrix}
       1&1\\
       1&1
       \end{pmatrix} 
       \;,\;
       E_4=\frac{1}{5}
       \begin{pmatrix}
       1&-\sqrt{2}e^{i\theta}\\
       -\sqrt{2}e^{-i\theta}&2
       \end{pmatrix}\;,\label{z2}
\end{align}
where $\theta\in\left(\frac\pi3,\frac\pi2\right]$ is determined by 
\be
    \cos(\theta)=\frac{1}{2\sqrt{2}}\;.
\ee
\end{example}

{\it Application.\;}SIC POVMs have been used to represent quantum states as points in a probability simplex~\cite{Appleby2011}.
In order to emulate these results in the formalism of semi-SIC POVMs, we will calculate the general form of the dual basis of a semi-SIC POVM in arbitrary finite dimensions and use this to represent the elements of our two dimensional semi-SIC POVMs in terms of probability vectors. 

First, we calculate the dual basis of a semi-SIC POVM in arbitrary finite dimensions. Let $\{E_x\}_{x=1}^{d^2}$ be a semi-SIC POVM and $\{F_y\}_{y=1}^{d^2}$ denote its dual basis. By definition, for all $y \in \{1,...,d^2\}$, the operators $F_y$  must satisfy the condition 
\be
\tr[E_xF_y]=\delta_{xy}\;.
\ee
It is straightforward to check that the matrices $\{F_y\}_{y=1}^{d^2}$ which satisfy the above relation are given by
\begin{align*}
    &F_y =\begin{cases}
    \frac{1}{a_{-}^{2}-b}E_y+\frac{a_{+}^{2}-a_{-}^2}{(a_{-}^{2}-b)(1-d)}S+\frac{1}{1-d}I_d  &\text{if }1\leq y\leq k\\
    \frac{1}{a_{+}^{2}-b}E_y+\frac{a_{-}^{2}-a_{+}^2}{(a_{+}^{2}-b)(1-d)}T+\frac{1}{1-d}I_d &\text{otherwise} 
    \end{cases}
\end{align*}
where 
$$S \eqdef \sum_{y=1}^k E_y \quad\text{and}\quad T \eqdef \sum_{y=k+1}^{d^2}E_y \; .$$

\begin{example}\label{example2}
As an explicit example, consider the qubit case where $b=\frac{2}{25}$. The dual basis operators are given by 
\ba\label{123}
    F_1&=\frac{25}{2}E_{1}-\frac{5}{2}(E_1+E_2)-I\;,\\
    F_2&=\frac{25}{2}E_{2}-\frac{5}{2}(E_1+E_2)-I\;,\\
    F_3&=\frac{25}{7}E_{3}+\frac{5}{7}(E_3+E_4)-I\;,\\
    F_4&=\frac{25}{7}E_{4}+\frac{5}{7}(E_3+E_4)-I\;.
\ea
For a 2-dimensional SIC POVM the relation is more symmetric, given as $F_x=6E_x-I_2$ for all $x\in\{1,2,3,4\}$. Note that, in the semi-SIC POVM, the coefficient $\frac72$ breaks the symmetry in the sense that $I=E_1+E_2+E_3+E_4$ is replaced by a weighted combination of $E_1+E_2$ and $E_3+E_4$.
\end{example}

The dual basis of the semi-SIC POVM can be used to derive a representation of a quantum state in terms of the probability distribution associated with the outcomes of the semi-SIC POVM. Specifically, since the dual basis $\{F_y\}_{y=1}^{4}$ as given in~\eqref{123} is a basis for the space of $2\times 2$ Hermitian matrices, we can express any $2\times 2$ density matrix $\rho$ as

\be 
    \rho =  \sum_{y=1}^{4} p_y F_y \;, \label{eq:7}
\ee
where  $p_y\eqdef\tr[E_y\rho]$ are probabilities associated with the semi-SIC measurement outcomes.

Furthermore, in general, not all probability vectors $\p=~(p_1,p_2,p_3,p_4)^T$ will yield in~\eqref{eq:7} a positive semi-definite matrix $\rho$. For example, $\vec{p}=(1, 0, 0, 0)$ is invalid because it results in $\rho=F_1$, which is a non-positive semi-definite matrix.

We now give an example in which we characterize the set of all such probability vectors $\p$ that give rise to a density matrix $\rho$ when $b=\frac2{25}$. 

\begin{example} Let $b = \frac2{25}$. Using the dual basis calculated in Example \ref{example2}, $\rho$ in Equation \eqref{eq:7} is positive semi-definite if and only if the polynomial
 \begin{align}
f(\p)\eqdef\det\left(\sum_{y=1}^{4} p_y F_y\right)\geq 0\;.
 \end{align}
 The polynomial $f$ can be calculated explicitly and is given by
 \ba \nonumber
 f(\p)=&-4p_1^2-4p_2^2-\frac{8}{7}p_3^2-\frac{8}{7}p_4^2+\frac{9}{2}p_1p_2+2p_1p_3\\
 &+2p_1p_4+2p_2p_3+2p_2p_4+\frac{9}{7}p_3p_4\;.
 \ea
Since $p_4=1-p_1-p_2-p_3$  the region $f(\p)\geq 0$ can be characterized by 
\ba\nonumber
&f(p_1,p_2,p_3,1-p_1-p_2-p_3)\\
=&-\frac{1}{14}(100p_1^2+100p_2^2+50p_3^2+25p_1p_2+50p_2p_3\\
 &+50p_1p_3-60p_1-60p_2-50p_3+16)\geq 0\;.
\ea 
This region, along with the region of a SIC-POVM, is plotted in Figure~\ref{figure1}, which displays the area of all $(p_1,p_2,p_3)$ with $p_1+p_2+p_3\leq 1$ that corresponds to quantum states. 

 \begin{figure}[h] 
 \includegraphics[scale=0.5]{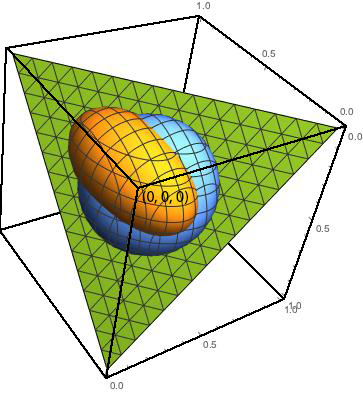}
 \caption{The probabilities which correspond to a quantum state. The green triangle is the surface with coordinates $p_1, p_2, p_3$ where $p_1+p_2+p_3=1$. The blue ellipsoid is the probability distribution associated with a SIC-POVM (i.e. $b=\frac1{12}$), while the yellow ellipsoid is the distribution associated with the semi-SIC POVM where $b=\frac{2}{25}$. Note that the areas of both distributions are located entirely above the green triangle.}
 \label{fig:probablity}
 \label{figure1}\end{figure}
 \end{example}

In the field of quantum state tomography, the Bloch-vector parametrization of a quantum state shows that SIC POVMs can be used to form an efficient quantum tomography~\cite{PR2012}. Additionally, it has been experimentally implemented in~\cite{REK2004, PhysRevX.5.041006}. Inspired by this, in the appendix we express the probabilities of obtaining a measurement outcome with respect to a semi-SIC POVM in terms of the Bloch vector representation.  We expect this to be a promising direction for future experiments.

{\it Conclusions. \;}In this paper, we demonstrated that in dimension two, the third condition in the definition of a SIC POVM (see Definition~\ref{def1}) does not follow trivially from the second condition. In particular, we established that without requiring this condition, the trace of any given element of the IC POVM can take (at most) the two distinct values given in~\eqref{eq:1}. This prompted us in Definition~\ref{main} to introduce a new class of measurements which we referred to as \emph{semi-SIC POVMs}. We then constructed the entire two dimensional one-parameter family of semi-SIC POVMs in Theorem~\ref{thm} (see an explicit example in Eqs.~(\ref{z1},\ref{z2})). We generalized several defining characteristics of semi-SIC POVMs in all finite dimensions, including the formulas for the values of $b$ (see \eqref{eq:3}) and  $k$ (see~\eqref{eq:4}). Finally, we showed that the dual basis of a semi-SIC POVM can be computed and is given by a simple formula similar to the formula of the dual of a SIC POVM. We then used it to represent any quantum state in terms of the probability vector associated with semi-SIC POVMs, analogous to the way it is done for SIC POVMs~\cite{Appleby2011}. 

Our construction of semi-SIC POVMs in two dimensions reveals a parametrized family of semi-SIC POVMs characterized by the parameter $b\in\left(\frac1{16},\frac1{12}\right]$. This continuous range of $b$ is in sharp contrast to the  discrete values of $b$ valid in dimension $d\geq 3$. In particular, Eq.~\eqref{eq:3} demonstrates that for $d\geq 3$, $b$ can take at most $d-1$ discrete values. Thus, it may still be the case that semi-SIC POVMs--which are not SIC POVMs-- do not exist in dimension $d\geq 3$. If this is the case, it would mean that the conditions in Definition~\ref{main} are sufficient to define a SIC POVM.  Whether semi-SIC POVMs--which are not SIC POVMs--exist in higher dimensions is left as an open question.

\begin{acknowledgments}
GG appreciate discussions on the subject with Dante Bencivenga, Taylor Kergan, and Gaurav Saxena.    
The authors acknowledge support from the Natural Sciences and Engineering Research Council of Canada (NSERC). KG acknowledge support from the Department of Mathematics and Statistics at the University of Calgary for an undergraduate research award.
\end{acknowledgments}

\bibliographystyle{apsrev4-1}
\bibliography{QRTbib}

\begin{appendix}

\section{Appendix}
Let $\rho$ be a density matrix for a qubit. Then it can be written in terms of the Bloch parameters $(r_x, r_y, r_z)$ as 
\begin{equation}
\rho=\frac{1}{2}\left(I_2+r_x\sigma_x+r_y\sigma_y+r_z\sigma_z\right)
\end{equation}
where $I_2$ is the $2\times 2$ identity matrix and $\sigma_x$, $\sigma_y$ and $\sigma_z$ are three Pauli matrices.
In the measurement of a semi-SIC POVM, by Born's rule, $q_i=\tr[E_i\rho]$ is the probability for obtaining the outcome $i\in[4]$.
The probability vector $\qq\eqdef(q_1,q_2,q_3,q_4)$ is in one-to-one correspondence with the Bloch vector $\rr\eqdef(r_x, r_y, r_z)$ via
\begin{align*}
q_1&=a_{-}(1+r_z)\\
q_2&=a_{-}\left[1+\left(2r\sqrt{1-r^2}\right)r_x+\left(2r^2-1\right)r_z\right]\\
q_3&=a_+\left[1-\left(\frac{2\sqrt{2}}{3}\cos\theta\right) r_x-\left(\frac{2\sqrt{2}}{3}\sin\theta\right) r_y-\frac{1}{3}r_z\right]\\
q_4&=a_+\left[1-\left(\frac{2\sqrt{2}}{3}\cos\theta\right) r_x+\left(\frac{2\sqrt{2}}{3}\sin\theta\right) r_y-\frac{1}{3}r_z\right]
\end{align*}
where $a_{\pm}$ , $b$ and $r$ are defined in the main text.

\end{appendix}

\end{document}